\documentclass[letterpaper,twocolumn,english,aps,prb,floatfix,showpacs,superscriptaddress,amssymb,preprintnumbers]{revtex4}
\usepackage{epsfig,latexsym,amssymb,amsmath,amsbsy,graphics,graphicx}
\usepackage{dcolumn,bm,amsfonts}
\usepackage{babel}
\usepackage{amsmath}
\usepackage{amsfonts}
\usepackage{amssymb}
\usepackage{graphicx}
\usepackage{amssymb}

\makeatletter

\providecommand{\tabularnewline}{\\}

\usepackage{amscd}
\usepackage{bm}

\usepackage{babel}
\makeatother
\begin{document}

\title{Quantum anisotropic Heisenberg chains with superlattice structure:
a DMRG study.}

\author{J. Silva-Valencia }

\affiliation{Departamento de F\'{\i}sica, Universidad Nacional de Colombia, A.
A. 5997, Bogot\'a, Colombia}

\affiliation{Instituto de F\'{\i}sica Gleb Wataghin, Unicamp, Caixa Postal 6165,
13083-970 Campinas SP, Brazil}

\author{J. C. Xavier}

\affiliation{Instituto de F\'{\i}sica Gleb Wataghin, Unicamp, Caixa Postal 6165,
13083-970 Campinas SP, Brazil}

\author{E. Miranda}

\affiliation{Instituto de F\'{\i}sica Gleb Wataghin, Unicamp, Caixa Postal 6165,
13083-970 Campinas SP, Brazil}

\date{\today{}}

\begin{abstract}
Using the density matrix renormalization group technique, we study
spin superlattices composed of a repeated pattern of two spin-1/2
XXZ chains with different anisotropy parameters. The magnetization
curve can exhibit two plateaus, a non trivial plateau with the magnetization
value given by the relative sizes of the sub-chains and another trivial
plateau with zero magnetization. We find good agreement of the value
and the width of the plateaus with the analytical results obtained
previously. In the gapless regions away from the plateaus, we compare
the finite-size spin gap with the predictions based on bosonization
and find reasonable agreement. These results confirm the validity
of the Tomonaga-Luttinger liquid superlattice description of these
systems.
\end{abstract}


\maketitle

\section{Introduction}
The synthesis of new materials which can be described in terms of
spin chains or spin ladders has revived the study of quantum spin
systems in one dimension in the last few years.\cite{dagottoandothers}
These systems are known to have surprising features such as the presence
of a gap in isotropic Heisenberg chains of integer spins and its absence
when the spins are half-integers.\cite{haldaneconj1,haldaneconj2,zimanschulz,takahashi,white,hallbergetal}
More recently, many experimental and theoretical results have shown
the presence of plateaus, in which the magnetization in an external
magnetic field is quantized to fractions of the saturated value.\cite{ajiroetal,shiramuraetal,cabraetal1,chenetal,citroetal,chenetal2}
These systems have special spatial structures, such as p-merization
or ladder geometry, which are responsible for the appearance of the
magnetization plateaus. Oshikawa, Yamanaka and Affleck\cite{oshikawaetal}
derived the condition $p\left(S-m^{z}\right)=\mathrm{integer}$, necessary
for the appearance of the magnetization plateaus in 1D systems. Here,
$p$ is the number of sites in the unit cell of the magnetic ground
state, $S$ is the magnitude of the spin and $m^{z}$ is the magnetization
per site (taken to be in the $z$-direction). The plateau state can
be viewed as a gapped state with nonzero magnetization, the integer
spin chains being a special case where $p=1$ and $m^{z}=0$.

Other types of spatial structures, such as quasi-periodic
couplings,\cite{arlegoetal} an inhomogeneous magnetic
field\cite{yamamotoetal2} or a superlattice structure\cite{valencia3}
can also give rise to magnetization plateaus.  The superlattice case
with periodic boundary conditions was studied by two of us in a
previous work.\cite{valencia3} There, we considered a spin
superlattice (SS) composed of a repeated pattern of two long and
different spin-$\frac{1}{2}$ XXZ chains. This model can be viewed as
the limit of $p$-merized chains when the number of sites per unit cell
is very large. Magnetization plateaus were found, with magnetization
values that depend on the relative sizes of the sub-chains, in
accordance with the condition of Ref.~\onlinecite{oshikawaetal}. The
determination of the width and magnetization values of the plateaus
relied on the Bethe Ansatz exact solution of the XXZ
chain.\cite{yangyang} The low-energy properties in the gapless regions
away from the plateaus, however, could be described by bosonization in
terms of a Tomonaga-Luttinger liquid superlattice
(TLLS).\cite{valencia12} This type of system is of potential interest
in nanoelectronic applications, where nanowire superlattice structures
have been built with semiconducting carbon
nanotubes.\cite{tllsnanotube} A TLLS would be obtained if the metallic
analogue could be manufactured.

\begin{figure}
\begin{center}\includegraphics[%
  width=3in,
  keepaspectratio]{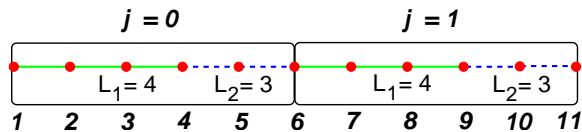}\end{center}

\caption{\label{cap:fig1} Spin superlattice structure with two unit cells
($N_{c}=2$), $L_{1}=4$, and $L_{2}=3$. The solid bonds correspond
to $\Delta_{1}$ and the dashed ones to $\Delta_{2}$. The total number
of sites is $L=\left(L_{1}+L_{2}-2\right)N_{c}+1=11$ (open boundary
conditions).}
\end{figure}

In this work, it is our purpose to check numerically the analytical
predictions based on the Bethe Ansatz and bosonization using the density
matrix renormalization group (DMRG).\cite{white,white2} In particular,
we calculate the magnetization curve, characterize its plateaus and
determine the effective Tomonaga-Luttinger liquid parameters for spin
superlattices. The unit cell of each SS consists of two $S=1/2$ XXZ
sub-chains with different anisotropy parameters $\Delta_{\lambda}$
and sizes $L_{\lambda}$ ($\lambda=1,2$) (but the same planar coupling),
as shown schematically in Fig.~\ref{cap:fig1}. The Hamiltonian of
the SS is written as a sum over $N_{c}$ unit cell Hamiltonians\begin{equation}
H_{SS}=\sum_{j=0}^{N_{c}-1}H_{c}\left(j,\Delta_{1},L_{1},\Delta_{2},L_{2}\right).\label{HSS}\end{equation}
Each unit cell, on the other hand, consists of a total of $L_{1}+L_{2}-2=L_{c}$
bonds. The first $L_{1}-1$ bonds have anisotropy parameter $\Delta_{1}$
and the following $L_{2}-1$ bonds have anisotropy parameter $\Delta_{2}$,
with Hamiltonian\begin{eqnarray*}
H_{c}\left(j,\Delta_{1},L_{1},\Delta_{2},L_{2}\right) & = & \sum_{n=1}^{L_{1}-1}H\left(jL_{c}+n,\Delta_{1}\right)\\
 & + & \sum_{n=1}^{L_{2}-1}H\left(jL_{c}+L_{1}+n-1,\Delta_{2}\right),\end{eqnarray*}
where\[
H\left(n,\Delta\right)=S_{n}^{x}S_{n+1}^{x}+S_{n}^{y}S_{n+1}^{y}+\Delta S_{n}^{z}S_{n+1}^{z}.\]
In the last expression, $S_{n}^{x}$, $S_{n}^{y}$ and $S_{n}^{z}$
are spin-$\frac{1}{2}$ operators at the $n$-th site. The total number of
sites in the SS is, for open boundary conditions, $L=\left(L_{1}+L_{2}-2\right)N_{c}+1$
and, for periodic boundary conditions,
$L=\left(L_{1}+L_{2}-2\right)N_{c}$.
We have set
the transverse coupling to 1 to set the energy scale. Since we will
be interested in increasing both $L_{1}$ and $L_{2}$ while keeping
their ratio fixed, we define $P$ as the greatest common divisor of
$L_{1}$ and $L_{2}$ and study the behavior of the system as $P$
grows. Note that each sub-chain contains $L_{\lambda}-1$ bonds and
$L_{\lambda}$ spins. The spins at the boundaries of the sub-chains,
however, should be viewed as belonging to either sub-chain (see Fig.~\ref{cap:fig1}).
Of course, in the limit of large $L_{\lambda}$ in which we are interested,
this ambiguity at the boundaries is immaterial. 
Finally, the total Hamiltonian has an external magnetic field $h$
applied along the anisotropy $z$-axis\begin{equation}
H_{T}=H_{SS}-\sum_{n=1}^{L}hS_{n}^{z}.\label{Htot}\end{equation}

Let us first recall some known results on the XXZ model, which is
the basic building block of the SS. Using the Bethe Ansatz, Yang and
Yang\cite{yangyang} found the exact
solution of the one-dimensional $S=1/2$ anisotropic (XXZ) Heisenberg
model. They showed that the model may exhibit three phases, according
to the value of the Ising anisotropy $\Delta$: a ferromagnetic (FM)
phase for $\Delta<-1$, a N\'{e}el antiferromagnetic (AFM) phase
with a spin gap for $\Delta>1$, and a gapless (critical) phase for
$-1<\Delta<1$. The low-energy properties of the gapless phase can
be described in terms of a Tomonaga-Luttinger liquid\cite{luttingerall}
with velocity $u$ and interaction parameter $K$.\cite{haldane1}

\section{Magnetization plateaus}

In the Hamiltonian (\ref{Htot}), the magnetic field couples to a
conserved quantity $S_{tot}^{z}=\sum_{n}S_{n}^{z}$. Thus, to obtain
the magnetization curve we only need the ground state energy at $h=0$,
$E(S_{tot}^{z},h=0)$, in each of the subspaces with fixed total spin
projection $S_{tot}^{z}\in\{0,1,\ldots,L/2\}$. Then, we can readily
obtain the energy in a finite magnetic field $h$ through the relation
$E(S_{tot}^{z},h)=E(S_{tot}^{z},0)-hS_{tot}^{z}$, from which we can
construct the magnetization curve.\cite{cabrareview} 

Since the DMRG is more precise and computationally faster with open
boundary conditions the magnetization curve was calculated this way.
We considered lattice sizes up to 160 sites keeping up to $m=150$
states per block. The discarded weight was kept around $10^{-12}$.
On the other hand, in order to compare with the analytical predictions
of the Tomonaga-Luttinger liquid parameters which were obtained with
periodic boundary conditions, we have also analyzed SS's with the
latter boundary conditions. In order to obtain a comparable accuracy,
we considered chains with up to 100 sites with up to $m=600$ states
per block. The truncation errors were below $10^{-9}$.

\begin{figure}
\begin{center}\includegraphics[%
  width=3in,
  keepaspectratio]{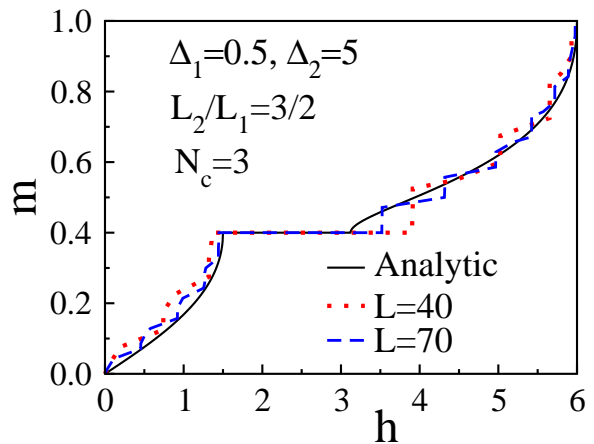}\end{center}

\caption{\label{cap:fig2}Magnetization density of a spin superlattice with
$\Delta_{1}=0.5$, $\Delta_{2}=5$, $L_{1}=2P$, $L_{2}=3P$ ($\ell=3/2$),
and $N_{c}=3$. The short-dashed line is for $P=3$ ($L=40$), and
the long-dashed one for $P=5$ ($L=70$). The solid line was obtained
analytically for a spin superlattice with long unit cells ($P\gg1$).\cite{valencia3}}
\end{figure}

Fig.~\ref{cap:fig2} shows the magnetization density $m$ (magnetization
per site normalized to the saturation value) of a SS as a function
of the external magnetic field for $\Delta_{1}=0.5$ and $\Delta_{2}=5$.
We set $N_{c}=3$, since we have observed that the magnetization curve
is fairly insensitive to the number of unit \textit{\emph{cells}}.
The sub-lattice sizes were chosen such that $L_{2}/L_{1}\equiv\ell=3/2$.
Two lattice sizes are shown in the figure: $L=40$ and $L=70$. They
correspond to $L_{1}=2P$ and $L_{2}=3P$, with $P=3$ and $P=5$,
respectively. For both sizes the magnetization density has a plateau
at $m=0.4$.

The continuous line in Fig.~\ref{cap:fig2} corresponds to the case
where we consider \emph{two long}  sub-lattices with a fixed size
ratio ($P\gg1$). In this case, the magnetization density of the SS
is given by\cite{valencia3}\begin{equation}
m=\frac{L_{1}m_{1}+L_{2}m_{2}}{L_{1}+L_{2}},\label{ms}\end{equation}
 where $m_{\lambda}$ is the magnetization per site of sub-chain $\lambda$.
The magnetization density of each sub-lattice was obtained from the
Bethe Ansatz solution in Ref.~\onlinecite{valencia3}.

It can be seen that the three curves have a magnetization plateau
at $m=0.4$. This has been shown to correspond to $m_{1}=1$ (FM phase)
and $m_{2}=0$ (AFM phase) in Eq.~(\ref{ms}).\cite{valencia3} The
plateau is a result of the ``spin incompressibility'' of both sub-chains:
sub-chain $1$ is magnetized at saturation, whereas sub-chain $2$
has a spin gap. However, the magnetic field width of the plateau is
dependent on the sub-chain sizes $L_{1}$ and $L_{2}$. In fact, the
fields at each end of the plateau are approached asymmetrically as
the system grows. The critical field $h_{c}$ (right-hand side of
the plateau) shows a larger finite size error than the saturation
field $h_{s}$ (left-hand side of the plateau).

\begin{table}
\begin{center}\begin{tabular}{cccc}
\hline \hline
$L$&
$h_{s}$&
$h_{c}$&
$\Gamma_{NT}$\tabularnewline
\hline 
10&
0.67719&
5.17056&
4.49336 \tabularnewline
40&
1.35468&
3.90474  &
2.55006 \tabularnewline
70&
1.44854&
3.51944&
2.07089\tabularnewline
100&
1.47166&
3.36001 &
1.88835 \tabularnewline
130&
1.48329&
3.27965&
1.79635\tabularnewline
160&
1.48900&
3.23373&
1.74473\tabularnewline
extrapolated&
1.495&
3.146 &
1.667\tabularnewline
prediction&
1.5&
3.121&
1.621\tabularnewline
\hline \hline
\end{tabular}\end{center}

\caption{\label{cap:tab1}The saturation field $h_{s}$, the critical field
$h_{c}$ and the width of plateau $\Gamma_{NT}$ at $m=0.4$ as a
function of lattice size. For all sizes $\Delta_{1}=0.5$, $\Delta_{2}=5$,
$L_{1}=2P$, $L_{2}=3P$ ($\ell=3/2$), and $N_{c}=3$. The second to last line is
the VBS extrapolation and the last one is the prediction of Ref.~\onlinecite{valencia3}.}
\end{table}

The behavior of the saturation field $h_{s}$, critical field $h_{c}$
and the width of the plateau $\Gamma_{NT}=h_{c}-h_{s}$ as a function
of the lattice size $L$ is shown in Table~\ref{cap:tab1}. Again
we focus on the plateau at $m=0.4$ and $\ell=3/2$ ($L_{1}=2P$,
$L_{2}=3P$, and $N_{c}=3$), $\Delta_{1}=0.5$, and $\Delta_{2}=5$.
We have found the thermodynamic limit of these quantities through
numerical extrapolation using the Vanden Broeck-Schwartz (VBS) algorithm,\cite{VBSall}
which is shown in the bottom line. The saturation field increases
with the lattice size, but slowly at large $L$, as seen in the first
column of Table~\ref{cap:tab1}. Extrapolation to infinite $L$ yields
the value $h_{s}=1.495$. As we have seen in connection with Eq.~(\ref{ms}),
we have $m_{1}=1$\textit{, i.e.,} the sub-lattice $1$ is totally
magnetized. As shown in Ref.~\onlinecite{valencia3}, the saturation
field is the field at which the corresponding homogeneous chain with
the same anisotropy parameter reaches saturation. For $\Delta_{1}=0.5$,
this happens at $h=1+\Delta_{1}=1.5$, very close to the numerical
value of $h_{s}$ and compatible with our interpretation.

In the second column of Table~\ref{cap:tab1}, it can be seen that
the critical field $h_{c}$ decreases with the lattice size. The convergence
is slower than for $h_{s}$ due to the smaller derivative of the magnetization
as a function of the field as the plateau is approached from above,
as is apparent in the analytical result of Fig.~\ref{cap:fig2}.
The extrapolated infinite size value of the critical field is $h_{c}=3.146$.
This should be compared with the gap of a homogeneous lattice with
anisotropy parameter $\Delta_{2}=5$, which is $h=3.121$. This value
is close to the numerically determined critical field. This agreement
is again consistent with the sub-lattice 2 with anisotropy parameter
$\Delta_{2}=5$ being in an AFM spin-gapped phase. The width of plateau
$\Gamma_{NT}=h_{c}-h_{s}$ as a function of $L$ is shown in the third
column of Table~\ref{cap:tab1}. The extrapolated infinite size limit
of the plateau width is $\Gamma_{NT}^{\infty}=1.667$.

\begin{figure}
\begin{center}\includegraphics[%
  width=3in,
  keepaspectratio]{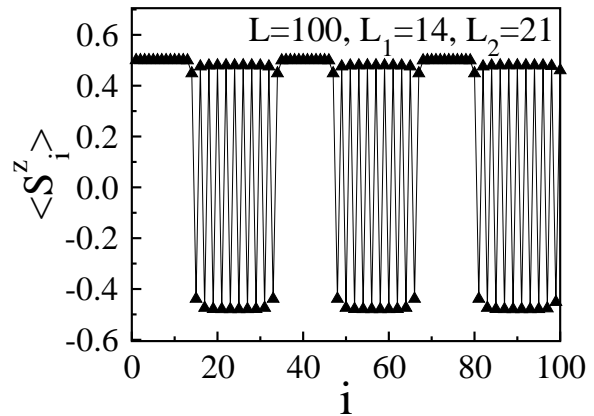}\end{center}

\caption{\label{cap:fig3} The magnetization profile of a spin superlattice
at the magnetization plateau of $m=0.4$. The parameters are $\Delta_{1}=0.5$,
$\Delta_{2}=5$, $L_{1}=14$, and $L_{2}=21$ ($\ell=3/2$, $L=100$).}
\end{figure}

In Fig.~\ref{cap:fig3}, the magnetization profile at the plateau
of $m=0.4$ is shown. The parameters used are $L=100$ ($L_{1}=2P$,
$L_{2}=3P$, $P=7$, and $N_{c}=3$), $\ell=3/2$, $\Delta_{1}=0.5$,
and $\Delta_{2}=5$. We can see that, indeed, in the sub-lattices
with anisotropy parameter $\Delta_{1}=0.5$, the spins are fully polarized,
whereas the sub-lattices with anisotropy parameter $\Delta_{2}=5$
are antiferromagnetically ordered.

The overall picture resulting from Figs.~\ref{cap:fig2} and \ref{cap:fig3}
and Table~\ref{cap:tab1} is thus compatible with the SS having a
nontrivial plateau at $m=1/(1+\ell)$, in which one sub-chain is fully
saturated with $m_{1}=1$, whereas the other is in a spin-gapped AFM
phase with $m_{2}=0$.\cite{valencia3,parity}%

\begin{figure}
\begin{center}\includegraphics[%
  width=3in,
  keepaspectratio]{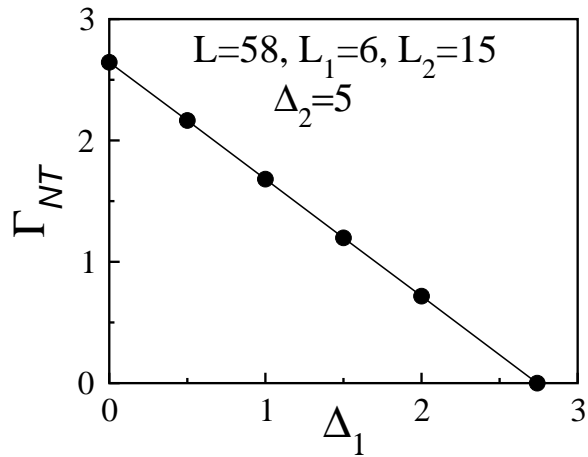}\end{center}

\caption{\label{cap:fig4} The width of nontrivial plateau $\Gamma_{NT}$
as a function $\Delta_{1}$. In this case $\Delta_{2}=5$, $L_{1}=6$,
$L_{2}=15$ ($\ell=5/2$, $m=\left(1+\ell\right)^{-1}=2/7$), $N_{c}=3$,
and $L=58$.}
\end{figure}

The width of the plateau of a SS with $\Delta_{2}=5$ is shown in
Fig.~\ref{cap:fig4} as a function of the anisotropy parameter $\Delta_{1}$.
The other parameters are $L_{1}=2P$, $L_{2}=5P$ ($\ell=5/2$, $m=\left(1+\ell\right)^{-1}=2/7$),
$P=3$, $N_{c}=3$, and $L=58$. We observe that the width of the
plateau decreases linearly with $\Delta_{1}$ and vanishes at $\Delta_{1}=2.741$.
The magnetization curve of this SS does not have plateaus for $\Delta_{1}>2.741$.
This linear dependence is expected. As we have seen, the plateau width
is given by $\Gamma_{NT}=h_{c}^{P}(\Delta_{2}=5)-1-\Delta_{1}$, where
$h_{c}^{P}$ is the gap of the corresponding homogeneous spin chain
in the AFM phase.\cite{valencia3} The slope of the line in the Fig.~\ref{cap:fig4}
is $-0.966$, close to the expected value of $-1$.

\begin{figure}
\begin{center}\includegraphics[%
  width=3in,
  keepaspectratio]{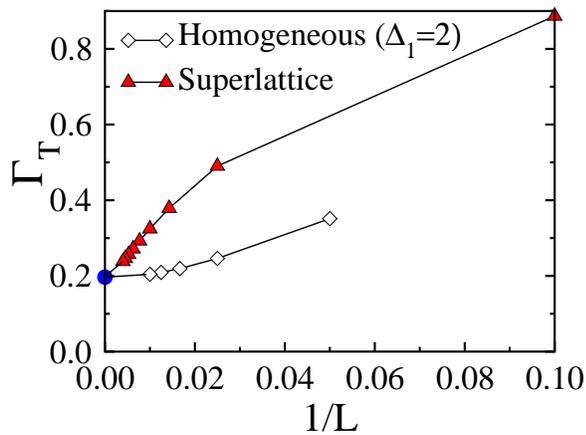}\end{center}

\caption{\label{cap:fig5} The width of the plateau at zero magnetization
as a function of the lattice size for a SS with $\Delta_{1}=2$, $\Delta_{2}=5$
and $\ell=3/2$ ($L_{1}=2P$, $L_{2}=3P$, and $N_{c}=3$) (closed
triangles). For comparison, we also show the size dependence of the
spin gap of a homogeneous chain with anisotropy parameter $\Delta=2$
(open diamonds). The extrapolated infinite size limits of the two
curves (closed circles) coincide within the numerical error.}
\end{figure}

The magnetization curve in Fig.~\ref{cap:fig2} has one nontrivial
plateau at $m=0.4$ for the parameters $\Delta_{1}=0.5$ and $\Delta_{2}=5$.
When both $\Delta_{1}>1$ and $\Delta_{2}>1$, however, a new plateau
at $m=0$ emerges. In this case, \emph{both} sub-chains 1 and 2 are
in a spin-gapped AFM phase.\cite{valencia3} For a SS with parameters
$\Delta_{1}=2$, $\Delta_{2}=5$, and $\ell=3/2$ ($L_{1}=2P$, $L_{2}=3P$,
and $N_{c}=3$), the width of the plateau at $m=0$ is shown in Fig.~\ref{cap:fig5}
as a function of $1/L$. Using the VBS algorithm, we determined the
infinite size limit of the plateau width to be $\Gamma_{T}=0.198$.
According to the analysis of Ref.~\onlinecite{valencia3}, this value
should be given by the gap of a homogeneous spin chain with anisotropy
parameter $\Delta_{1}=2$. For comparison, we also show in Fig.~\ref{cap:fig5}
the value of the latter gap as a function of system size. The extrapolated
gap size is $\Gamma_{T}=0.196$, while the value obtained from the
exact solution is $\Gamma_{T}=0.19842$. Indeed, the extrapolated
values coincide within the numerical error.

\section{Gapless region: The effective Tomonaga-Luttinger parameters}

The low-energy properties of the SS \emph{away from the plateaus}
can be described in terms of a TLLS.\cite{valencia12} The
Hamiltonian of TLLS is given by 

\begin{equation}
H_{LLSL}=\frac{1}{2\pi}\int dx\left\{ u(x)K(x)\left(\partial_{x}\Theta\right)^{2}+\frac{u(x)}{K(x)}\left(\partial_{x}\Phi\right)^{2}\right\} ,\label{HLLS}\end{equation}
 where $\partial_{x}\Theta$ is the momentum field conjugate to $\Phi$:
$[\Phi(x),\partial_{y}\Theta(y)]=i\delta(x-y)$. \textbf{}The fields
$\Phi$ and $\Theta$ can be related to the spin density operators.\cite{valencia3}
In the Hamiltonian (\ref{HLLS}), we have introduced the sub-chain-dependent
parameters $u(x)$ and $K(x)$. For $x$ in the sub-chain $\lambda$,
one has $K(x)=K(J,\Delta_{\lambda},h)$ and $u(x)=u(J,\Delta_{\lambda},h)$\textit{,
i.e.}, the usual uniform Tomonaga-Luttinger parameters for each sub-chain,
which can be obtained directly from the Bethe Ansatz solution.\cite{haldane1,cabraetal1}
Of course, this effective low-energy description is valid asymptotically
in the limit of very long sub-chains. Using periodic boundary conditions
and diagonalizing the Hamiltonian (\ref{HLLS}), we find that the
low energy properties of the SS are determined by just a few effective
parameters.\cite{valencia3,valencia12} These parameters
are the effective velocity $c$ and the effective correlation exponents
$K^{\ast}$ and $\overline{K}$, which are given by\cite{valencia3,valencia12}
\begin{equation}
c=\frac{u_{1}(1+\ell)}{\sqrt{1+\eta\ell u_{1}/u_{2}+\left(\ell u_{1}/u_{2}\right)^{2}}},\label{velo}\end{equation}

\begin{equation}
K^{\ast}=\frac{\sqrt{1+\eta\ell u_{1}/u_{2}+\left(\ell u_{1}/u_{2}\right)^{2}}}{\frac{1}{K_{1}}+\ell\frac{1}{K_{2}}\frac{u_{1}}{u_{2}}}\equiv f(K_{1},K_{2}),\label{kstar}\end{equation}
\begin{equation}
\overline{K}=f(1/K_{1},1/K_{2}),\label{kover}\end{equation}
 where $\eta=K_{1}/K_{2}+K_{2}/K_{1}$. Clearly, $(c,K^{\ast},\overline{K})\rightarrow(u_{2},K_{2},1/K_{2})$
as $\ell\rightarrow\infty$, and $(c,K^{\ast},\overline{K})\rightarrow(u_{1},K_{1},1/K_{1})$
as $\ell\rightarrow0$, as expected. The important feature to notice
in Eqs.~(\ref{velo}), (\ref{kstar}), and (\ref{kover}) is the
fact that the effective SS parameters represent a certain weighted
average of the individual sub-chain velocities and correlation exponents.
This weighted average is induced by the superlattice structure and
is a feature ubiquitous in TLLS's.\cite{valencia3,valencia12}

\begin{figure}
\noindent \begin{center}\includegraphics[%
  width=3in,
  keepaspectratio]{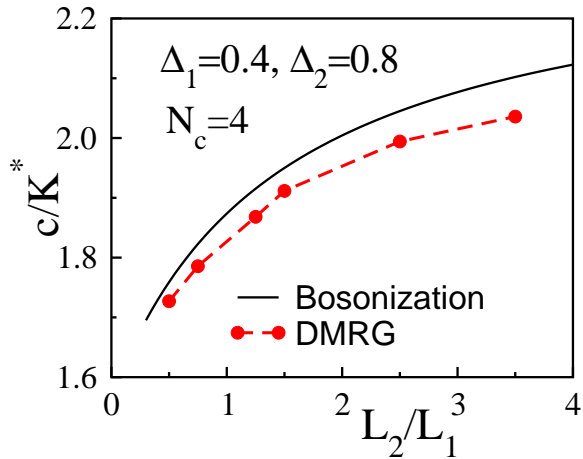}\end{center}

\caption{\label{cap:fig6} $c/K^{\ast}$ as a function $\ell$ for a SS with
$\Delta_{1}=0.4$ and $\Delta_{2}=0.8$. The continuous line was obtained
using bosonization.\cite{valencia3} The closed dots correspond to
the DMRG results.}
\end{figure}

It is straightforward to extract from the Hamiltonian (\ref{HLLS})
the finite-size spin gap of the system. It is given by

\begin{equation}
E\left(S_{tot}^{z}=1,h=0\right)-E\left(S_{tot}^{z}=0,h=0\right)=\frac{\pi c}{2K^{\ast}L}.\label{cke}\end{equation}
Thus, from the scaling of the spin gap with the system size, we can
verify the predictions of Eqs.~(\ref{velo}) and (\ref{kstar}) for
the effective Tomonaga-Luttinger parameters.

Fig.~\ref{cap:fig6} shows the numerically determined ratio $c/K^{\ast}$
for a SS with $\Delta_{1}=0.4$ and $\Delta_{2}=0.8$ as a function
of $\ell$. We used $N_{c}=4$ and $L=100,76,100,52,76,100$, for
$\ell=1/2,3/4,5/4,3/2,5/2,7/2,$ respectively. For comparison, we
also show the TLLS prediction obtained from the ratio of Eqs.~(\ref{velo})
and (\ref{kstar}) and from the known values of $u_{\lambda}$ and
$K_{\lambda}$ for homogeneous chains. We can see that there is reasonable
agreement, with slightly larger discrepancies at larger $\ell$. The
ratio $u/K$ for homogeneous chains with anisotropy parameters $\Delta_{1}=0.4$
and $\Delta_{2}=0.8$ are equal to $u_{1}/K_{1}=1.57$ and $u_{2}/K_{2}=2.33$,
respectively. $c/K^{\ast}$ interpolates smoothly between $u_{1}/K_{1}$
and $u_{2}/K_{2}$ as $\ell$ increases, a manifestation of the spatial
averaging due to the superlattice structure. We believe the small
discrepancies between the curves in Fig.~\ref{cap:fig6} are due
to the finite sizes of the sub-chains. We recall that the TLLS predictions
are expected to hold asymptotically for very long sub-chains. For
a gapless phase, the inhomogeneities created by the boundaries between
sub-chains will give rise to Friedel oscillations which die out only
as power laws.\cite{eggergrabert} These disturbances are expected
to give rise to finite-size corrections to the TLLS predictions. We
stress, however, that although the TLLS analysis predicts a sort of
weighted average for the dependence of $c/K^{\ast}$ on $\ell$, the
detailed form of this average is highly nontrivial. Yet, precisely
this non-linear dependence is strikingly confirmed by the numerical
data. We consider this as a stringent test of the predictions of the
theory.

\section{Conclusions}

In summary, we have used the finite size DMRG method with open and
periodic boundary conditions to study spin superlattices made up of
a periodic arrangement of two XXZ chains with different parameters
and sizes. We confirmed previous analytical predictions of a nontrivial
plateau in the magnetization curve at $m=1/(1+\ell)$, where $\ell$
is the relative size of the sub-chains. When both anisotropies are
larger than $1$, we have also confirmed the expected trivial plateau
at $m=0$. The nontrivial plateau width was shown to approach the
asymptotic value of a superlattice with long sub-chains. Moreover,
the magnetization profile was seen to be in accord with the analytical
predictions of one sub-chain being saturated and the other one being
AFM ordered. Finally, we found fairly good agreement in the gapless
region with the results of a Tomonaga-Luttinger liquid theory as applied
to a superlattice structure. This agreement confirms the non-trivial
prediction of this theory for the way the individual sub-chain properties
are averaged over in the effective low-energy description of the superlattice.

\begin{acknowledgments}
J. Silva-Valencia is grateful to A. L. Malvezzi for some help with
the DMRG method. This work was supported by FAPESP through grants
01/07778-0 (J. S.-V.), 00/02802-7 (J. C. X.) and 01/00719-8
(E. M.), and by CNPq through grant 301222/97-5 (E. M.).

\end{acknowledgments}


\begin{thebibliography}{39}
\bibitem{parity}Because of the small sizes, we have kept the AFM sub-chains with an
odd number of sites, such that no frustration of the AFM order is
introduced. This is expected to be irrelevant for long sub-chains.
\end{thebibliography}
\end{document}